\documentstyle[12pt,aasms4]{article}

\begin{document}

\title{Spectral Evolution of PKS 2155-304 observed with $BeppoSAX$
during an Active Gamma-ray Phase}

\author{L.Chiappetti \altaffilmark{1} 
\authoraddr{Lucio Chiappetti , IFCTR, via Bassini 15, I-20133 Milano, Italy, e-mail
lucio@ifctr.mi.cnr.it}, 
L.Maraschi \altaffilmark{2}, 
F.Tavecchio \altaffilmark{2,1},
A.Celotti \altaffilmark{3}, 
G.Fossati \altaffilmark{3}, 
G.Ghisellini \altaffilmark{2}, 
P.Giommi \altaffilmark{4}, 
E.Pian \altaffilmark{5},
G.Tagliaferri \altaffilmark{2}, 
A.Treves \altaffilmark{6}, 
C.M. Urry \altaffilmark{7}, 
Y.H. Zhang \altaffilmark{3}}

\altaffiltext{1}{Istituto di Fisica Cosmica G.Occhialini, IFCTR/CNR, Milano, Italy}
\altaffiltext{2}{Osservatorio Astronomico di Brera, Milano and Merate, Italy}
\altaffiltext{3}{International School of Advanced Studies, SISSA/ISAS, Trieste, Italy}
\altaffiltext{4}{BeppoSAX Science Data Centre, SDC/ASI, Rome, Italy}
\altaffiltext{5}{Istituto ITESRE, CNR, Bologna, Italy}
\altaffiltext{6}{Istituto di Fisica, Universit\`a dell'Insubria, Como, Italy}
\altaffiltext{7}{Space Telescope Science Institute, Baltimore MD, USA}

\slugcomment{Submitted to ApJ 19/11/1998 Revised 11/03/1999 Accepted 19/03/1999}

\begin{abstract}
We present the results of $BeppoSAX$ observations of PKS~2155--304
during an intense $\gamma$--ray flare. The source was in a high
X--ray state.  A temporal analysis of the data reveals a tendency of
the amplitude of variations to increase with energy, and
the presence of a soft lag with a timescale of the
order $10^3$ s.  A curved continuum spectrum, with no
evidence of spectral features, extends up to $\sim$ 50 keV, while
there is indication of a flatter component emerging at higher energies,
consistent with the interpretation of the broad band spectral energy
distribution (SED) as due to synchrotron self--Compton (SSC) emission
from a single region. Notably, the fitting of the SED with such a
model is consistent with an interpretation of the detected soft lag
as due to radiative cooling, supporting the
idea that radiation losses play an important role in variability.
The observed shifts of the SED peaks between the lowest and highest flux levels
can be accounted for by an increase of the ``break'' energy
 in the relativistic particle spectrum.
The model predicts emission at TeV energies in good agreement with the
recently reported detection.
\end{abstract}


\section{Introduction}

BL Lacertae objects are a rare class of Active Galactic Nuclei,
characterized by strong and variable non-thermal emission, extending
from the radio to the gamma-ray band. The non-thermal continuum is
commonly attributed to synchrotron and inverse Compton (IC) radiation
emitted in a relativistic jet pointing towards the observer (e.g.Urry \&
Padovani 1995).
                 
PKS 2155-304 is one of the brightest BL Lac in the X-ray band and one
of the few detected in $\gamma$-rays by the EGRET experiment on CGRO
(Vestrand, Stacy \& Sreekumar 1995). Its broad band spectrum indicates 
that the
radio to X-ray emission is due to synchrotron radiation with a peak in
the power per decade distribution between the UV and the soft X-ray
band, corresponding to the definition of an High Frequency Peak BL Lac
(HBL) (Padovani \& Giommi 1995). The gamma-ray spectrum in the EGRET
range (0.1--10 GeV) is flat ($\alpha_{\gamma}\simeq 0.7$) indicating
that the peak of the inverse Compton power is beyond $\sim$ 10 GeV. 
The source has only recently been detected in the TeV band  (see
below).

In the past years PKS 2155-304 has been the target of numerous
multiwavelength campaigns involving observations from the X-rays to
longer wavelengths.  In the May 1994 campaign (Urry et al. 1997) a
well defined flare was seen in the X-ray band (ASCA), followed
by less pronounced flares in the EUVE and UV ranges, lagging the X-rays
by one and two days, respectively. A time lag between variations
in different X-ray bands (0.5-1 and 2.2-8 keV) was also reported
(Makino et al. 1996).  In a previous multiwavelength campaign, based
on ROSAT and IUE, correlated low amplitude fluctuations were observed
but no lag larger than a few hours was seen between the X-ray and UV
variations (Edelson et al. 1995). 

According to the basic scenario the
X-ray emission is due to synchrotron radiation from the highest energy
electrons and the complex spectral variability observed in this band 
therefore
reflects the injection and radiative evolution of freshly accelerated
particles.

No observations at other wavelengths simultaneous with one in gamma-rays
 were ever obtained previously for this source, yet it is essential to measure
the IC and synchrotron peaks at the same time, in order to
unambiguously constrain emission models (e.g Dermer, Sturner \&
Schlickeiser 1997; Tavecchio Maraschi \& Ghisellini 1998).

For this reason, having been informed by the EGRET team of their
observing plan and of the positive results of the first days of the
CGRO observation, we decided -- with the agreement of the $BeppoSAX$ TAC and
the collaboration of the $BeppoSAX$ SDC -- to swap a pre-scheduled target of
our $BeppoSAX$ blazar program with PKS 2155-304. During November 11-17
1997 (Sreekumar \& Vestrand 1997) the $\gamma$-ray flux from PKS
2155-304 was very high, roughly a factor of three greater than the
highest flux previously measured from this object. $BeppoSAX$ pointed at
this source for about 1.5 days starting November 22. A quick-look analysis
indicated that also the X-ray flux was close to the highest detected level 
 (Chiappetti \& Torroni 1997) and higher by a factor two than that 
observed by $BeppoSAX$ in 1996 (Giommi et al. 1998). 
During the completion of this
work we have been informed that the source was detected at
TeV energies by the University of Durham TeV telescope Mark 6
(Chadwick et al. 1998) at the time of the $BeppoSAX$ observations.

Here we report and discuss the data obtained by $BeppoSAX$.  The
structure of the paper is as follows: the relevant information on the
observations is given in section 2 and the data analysis methods are
presented in detail for each instrument in section 3; light curves and
temporal analysis are reported in section 4, while the spectral
results are the subject of section 5.  Finally, in section 6 the
implications for theoretical models are discussed. Conclusions are summarized
 in Section 7.

\section{Observations}

The $BeppoSAX$ scientific payload (see Boella et al. 1997a) includes
four Narrow Field Instruments (NFIs) pointing in the same
direction. Namely, there are two imaging instruments : the Low Energy
Concentrator Spectrometer (LECS), sensitive in the range 0.1--10 keV
(Parmar et al. 1997), and the Medium Energy Concentrator Spectrometer
(MECS), sensitive in the range 1.3-10 keV (Boella et al. 1997b) and
consisting of three identical units. Both the LECS detector and the
three MECS detectors are Gas Scintillation Proportional Counters
(GSPC) and are in the focus of four identical X-ray telescopes.
Additionally there are two collimated instruments: the High Pressure
Gas Scintillation Proportional Counter (HPGSPC),
sensitive in the 4-120 keV range (Manzo et
al. 1997) and the Phoswich Detector System (PDS), 
sensitive from 12 to 300 keV (Frontera et al. 1997).

$BeppoSAX$ NFIs observed PKS 2155-304 for slightly less than 1.5 days
from 16:03 UT of November 22 to 01:35 UT of November 24 1997.
The total exposure
time was of about 22 ks for the LECS and 63 ks for the two MECS units
now in operation (M2 and M3),
with an average count rate of 0.64 (1.8-10 keV)
cts/s in a single MECS unit (corresponding to about 5 mCrab) and 2.21
cts/s in the LECS (0.1-4 keV band).

The HPGSPC and PDS were operated in the customary collimator rocking
mode, where each collimator points alternately at the source and at
the background for 96 s. In the case of the HPGSPC there is a single
collimator, so the whole instrument is looking at the target for half of the
time, and in the offset negative direction for the other half. The resulting
source exposure time was 29 ks and the source was detected at least up to 30
keV with a rate of 0.5 cts/s. The PDS has two collimators: at any time
one of them looks at the source, and the other one at the background.
Therefore the target is observed continuously (in our case the
exposure time was 28+29= 57 ks) but using only half of the collecting
area, with a count rate between 0.2 and 0.3 cts/s.

\section{Data Reduction}

The data reduction for MECS, PDS and HPGSPC was done using the XAS
software (Chiappetti \& Dal Fiume 1997) on telemetry files contained in
the Final Observation Tapes.  For the LECS we have used the linearized,
cleaned event files version rev.1.0 generated at the $BeppoSAX$ Science Data Centre (SDC).

We have preliminarily selected the intervals in which the satellite was
pointing at the source (unocculted by the Earth,
i.e. Earth elevation angle greater than 3 degrees)
using the information in the attitude files. 
In addition, for collimated
instruments, we have excluded the first 5 minutes after egress from the
South Atlantic Geomagnetic Anomaly, when the instrument gain is being calibrated.

\subsection{Imaging instruments}

As part of the standard data accumulation for both LECS and MECS, a
number of corrections is applied to each photon, namely the positional
coordinates are linearized (correcting for geometrical distortion), the
energy of each photon is referred to the detector centre (compensating
for known spatial disuniformities of gain) and corrected to a standard
gain using the gain histories to remove the temperature dependency of
gain. In our case we verified the stability of the gain within at worst
2\% during the initial orbits.

For both instruments the preferred background subtraction method is to
use a background spectrum accumulated from blank field exposures in the
{\it same} detector region where the target spectrum is accumulated. The
accumulation of a simultaneous background spectrum in a {\it surrounding}
annulus would instead include a residual contribution from the target. In
fact, an extraction radius of the order of 8 arcmin, as used here,  leaves out
$\sim$ 2-4 \% of the PSF, depending on energy.

\subsubsection{MECS}

Source spectra and light curves were accumulated in a circular
region of 8.4 arcmin radius around the target position. In order to
improve particle background rejection, we also applied 
Burst
Length thresholds (channel 25-55 for M2 and 27-60 for M3) 
consistent with the standard response matrices.

The background was accumulated in the same region from a large dataset of
blank fields obtained during the $BeppoSAX$ Performance Verification (PV) phase
(and available to IFCTR as a $BeppoSAX$ hardware institute) for a total
exposure time of 1120 ks. In addition to the considerations made above,
in the case of the MECS this has to be preferred also because the
outermost part of the field of view is known to be affected by residual
contamination from misplaced calibration events (Chiappetti et al. 1998).

\subsubsection{LECS}

The LECS data analysis was based on the cleaned and linearized event
files processed at the SDC. The events were extracted with XSELECT within
a circle centered on the source of radius of 8.5 arcmin. Since the count
rate of the source is very high, the background is not very important
and we used the standard background files supplied by SDC. The
spectral analysis was performed using the calibrations released in September 1997.

\subsection{Collimated instruments}

\subsubsection{PDS}

The standard PDS spectra accumulation occurs via the following procedure.
One first creates the time profiles of both collimator positions
(
sampled at 1 s resolution), and generates for each three sets
of time windows, corresponding to stable positions on source and offset
in the positive (off+) and negative (off--) directions. Then three
spectra in raw PHA channels for each unit are accumulated over the
respective time windows (background+source, and background in the off+ and off-- cases).

A merged background spectrum for each unit (
exposure-weighted sum
of the off+ and off-- spectra) is then generated and subtracted from
the on source spectrum.
In order to sum the
net spectra of the four units, one has to equalize the respective gain
(i.e. convert from PHA channels to keV using unit-specific relations
assuming a standard gain), which is done contextually with a
user-selected rebinning (usually logarithmic) in energy space.

During accumulation one also applies default rise time (RT) threshold
(channels 3-150) to improve rejection of non-X-ray events. 
For medium-weak sources  one can apply an additional set of narrower
energy-dependent RT thresholds (so called ``PSA correction'' method).
The RT of individual photons is compensated for temperature variations
with respect to the nominal temperature for which the thresholds have
been generated.
In this way a 40
\% reduction of the background is achieved (from 10 to 6 cts/s), with 
a smaller effect on the net rate (which is accounted for by a 10-20 \%
adjustment of the cross normalization of the PDS response matrix with
respect to the MECS one): in our case we obtain on the overall energy range 
$0.27\pm 0.05$ cts/s (no correction) and $0.21\pm 0.03$
cts/s (with correction), which are consistent within the errors.

We verified the quality of the background subtraction, by subtracting the
off+ and off-- raw background spectra of each unit from each other, and
then summing the difference spectra of the four units 
exactly in the same way described above for
the production of net spectra. Indeed the resulting signal ($0.06\pm 0.06$
cts/s) is consistent with zero.

PDS data may be affected by spurious spikes, due to the
{\it fluctuations} of meta-stable phosphorescence induced in the crystals by
particles, 
interpreted by the electronics
as spikes in the count rate (as high as 300 cts in a 0.1 s interval). 
A spike filtering has been applied to all
spectra and light curves,
rejecting all time intervals where the
scientific data rate (in the PHA range 10-350 and in the RT range
3-130) are above 25 cts/s.

Since the signal is weak, we have not attempted to accumulate binned
light curves. Similarly to MECS and HPGSPC, we used instead large bins
of variable size (of the order of 1 hr or less) corresponding to
unocculted intervals in each orbit, we accumulated three separate light
curves for each PDS unit (for the on source, off+ and off-- cases, taking in due account
the exposure fraction of each bin), made a weighted
combination of the off+ and off-- curves, subtracted it from the
source+background one, and finally summed the four unit net curves.

\subsubsection{HPGSPC}

The standard HPGSPC spectra accumulation is similar to the procedure
described above for the PDS. However one has only one collimator
position time profile and just two disjoint time windows, corresponding
to stable positions on source and offset in the negative directions.
Then one accumulates two spectra over the respective time windows
(background+source, and offset background), also applying the Burst
Length thresholds (channels 80-115) to improve rejection of non-X-ray
events.  The background has to be corrected 
before subtraction.
by adding a difference
spectrum which compensates for effects due to the different position of
the collimators (mainly the illumination by the calibration sources).

As for the PDS, we used ``coarse variable bins'', as defined above, to
generate 
light curves for the on source and offset cases.
We subtracted
the uncorrected background plus a further constant, corresponding to the
integral of the difference spectrum in the wished PHA channel range. We
intended to ``hook up'' our HPGSPC data with MECS 
and
PDS data 
in the overlapping energy ranges. 
However the HPGSPC 5-10 keV curve shows excesses with
respect to the MECS in a couple of orbits, related to
$^{55}Fe$
calibration events not
rejected properly by the onboard tagging. 
Further occasional deviations still present in the 7-10 keV range suggest
that the stability of the difference spectra with time is poorly known,
throwing doubts also on the confidence of HPGSPC 10-16 keV light curves.

\section{Results}

\subsection{Light curves}

In Fig. \ref{uno} we show the MECS light curve at
120 sec resolution. 
Epochs of high, low and intermediate intensity, which  will be used to
integrate spectra at different intensity levels, are indicated with 
letters A-F (see Section 4.2).
In Fig. \ref{due} (top panels) we show the light curves binned over 1000 s in
different energy bands: 0.1-2 keV for the LECS, 2-4 keV and 4-10 keV
for the MECS.
The source reaches a peak after the first 2
hours of observation (indicated as A in Fig. \ref{uno})
and declines by a factor 3 (max/min) in the
following 10 hours. A smaller flare (indicated as D) follows 
and the light curve seems
to stabilize in the interval indicated as F. 
The variability timescale appears to be well resolved
and no episode of very fast variability ($F dt/dF < 1 h$) is apparent.

The most
rapid variation observed (the decline from the peak at the start of
the observation) has a halving timescale of about $2\times 10^{4}$ s,
similar to previous occasions (see e.g. Urry et al. 1997).  Light
curves from the higher energy instruments were derived as detailed in
section 2, but no significant variability was detected due to poor
statistics.

The variability amplitude is different in the three bands, increasing
with increasing energy. In order to better characterize this energy
dependence, hardness ratios between the 2-4/0.1-2 keV (HR1) and the
4-10/2-4 keV bands (HR2) were computed. They are shown in the last
panels of Fig. \ref{due} (lower panels). HR1 has smaller uncertainties and
clearly increases with intensity in a correlated fashion on the
timescale of each peak.  However the correlation is not biunivocal:
HR1 has the same value at the first two peaks (A and D), which have
significantly different intensities, and is lower towards the end of
the observation, when the average intensity is similar to that of peak
D.  Note also that for both the first and the second peak the hardness
ratio is high already before the peak is reached.  HR2 does not show a
 clear trend, which could be due  to the larger
uncertainties and to the smaller energy range. The issue of spectral variability
is further discussed below (sect. 4.2.3).

We looked for time lags between variations at different energies as
suggested by the ASCA observations of this same source and of Mrk 421
(Makino et al. 1996, Takahashi et al. 1996). 
The presence of a soft lag is indicated by the
fact that the hardness ratio increases {\it before} the 
 intensity peaks, as it is the case in our light curve.

To quantify the lag we used the Discrete Correlation Function (DCF) method
developed by Edelson \& Krolik (1988) for data sets with irregular
spacings. We binned the light curves in smaller and smaller bins since no
lag was apparent for bin sizes larger than 1000 s. In Fig. \ref{tre} we
show the DCF obtained correlating the light curves in the bands 0.1-1.5
(LECS) and 3.5-10 keV (MECS). 
A gaussian fit, which takes into account the overall
symmetry of the distribution around the peak, yields a maximum
corresponding to a soft lag of 0.50 hr (with a 1$\sigma$ error of 0.08 hr).  We
also applied the Minimum Mean Deviation (MMD) method (Hufnagel \& Bregman
1992) (Fig. \ref{quattro}).  Here the correlation estimator is the mean
deviation of the two cross correlated light curves: the minimum mean
deviation corresponds to the best cross correlation.  A gaussian fit to
this minimum gives again a soft lag of 0.33 $\pm$ 0.07 hr. 

The issue of the lags and their
uncertainties, estimated with a Monte Carlo procedure, is discussed
in detail in Treves et al. (1998) and Zhang et al. (1999), where
a comparison is made with the 1996 $BeppoSAX$ (Giommi et
al. 1998) and 1994 ASCA data (Makino et al. 1996)

\subsection{Spectra}

We analyzed first the overall spectrum from the entire observation
separately for each instrument and then combining different
instruments.  We also studied spectral variability, accumulating
spectra in the time intervals indicated as A to F in Fig. \ref{uno}.
When not differently stated we used in the fits the value of the
galactic column density reported by Lockman \& Savage (1995), $
N_{H} =1.36\times 10^{20}$ cm$^{-2}$.

For the spectral fitting procedures the LECS and MECS data were binned
in energy according to the template provided by SDC (Fiore \& Guainazzi 1997) 
which takes into account the effective energy resolution of the
spectrometers.  For the high energy instrument data have been severely
rebinned as described below.  In all cases we considered for the LECS
the energy range 0.1-4 keV, for the MECS 2-10 keV.

The response matrices are generated according to the accumulation
conditions (extraction radius, binning etc.) for the MECS and PDS case
using the latest version of the XAS software, while for LECS and
HPGSPC case standard matrices are used, as released in September 1997
(with optional rebinning).
Results of the fits are summarized in Table 1.

\subsubsection{Single Instruments}

\paragraph{LECS}

A single power law is a poor fit to the LECS data, although it yields an
$N_{H}$ value quite close to the galactic one. An acceptable fit is
obtained by modelling the spectrum with a broken power law (see Table 1);
in the latter case too the $N_{H}$ determined by the fit is consistent
with the galactic one, which was therefore fixed in all the
following fits. The residuals of this fit do not show clear evidence of
spectral features in absorption or emission. In particular we do not
detect absorption near 0.6 keV as seen in some previous
observations (Canizares \& Kruper 1984; Madejski et al. 1991;
see however Brinkmann et al. 1994). 

\paragraph{MECS}

The MECS alone is not sensitive to low values of $N_H$ because of the
low energy cutoff induced by the Beryllium window. A fit with a single
power law and free column density gives $\chi ^2$ close to unity, but
yields an ill-determined $N_{H}$ of several $ 10^{21}$ cm$^{-2}$,
inconsistent with the results from the LECS. When fixing $N_{H}$ to
the galactic value, a broken power law model describes well the spectrum
within the 2-10 keV band (see Table 1). Note that the spectral index
in the higher energy range of the LECS (1-4 keV) is similar to that in
the lower energy range of the MECS (2-3 keV) giving us confidence that
the increasing slope suggested by the fits is a good representation of
the spectrum.

\paragraph{HPGSPC}

The HPGSPC data were rebinned ``ad hoc'' into 8 bins. The signal is
present very clearly ($>$ 4$\sigma$) up to 13 keV and with a lower
statistical significance up to at least 24 keV (i.e.  below the Xe K
edge).  As MECS and HPGSPC are very well cross-calibrated (Cusumano 
et al. 1998) and since this spectrum lies
well on the extrapolation of the MECS fit, it has been fitted only in
combination with spectra from other instruments.

\paragraph{PDS}

For the PDS we  used a very coarse grouping in just 4 logarithmic
bins. The signal from the source is clearly visible in the first bin
(up to 26 keV), and remains present at a non-zero level (although with
a poorer significance, $\simeq 2 \sigma$) in the other bins.  Fitting
the PDS data with a single power law yields a poor result, with
a spectral index {\it flatter} than in the MECS band.

\subsubsection{Combined Spectra}

We  fitted the composite LECS+MECS spectrum over the entire range
0.1-10.5 keV (see Table 1). Given that the single instrument spectra are well
described by  broken power law models, we tried to reproduce the
composite spectrum with the same model. We allowed for a free relative
normalization between the LECS and MECS data and the best fit value
found (LECS norm/ MECS norm$\simeq 0.7$ ) is consistent with previous
works. However the fit is unacceptable.
In Fig. \ref{cinque} we show the results of the fitting procedure :
 the model
does not represent the data well ($\chi_{red}^{2} =2.3$) and the shape of
the residuals indicates that the spectrum is flatter than the model at
the lowest energies and steeper at the highest ones. Thus a model
describing a more continuous steepening is required, in agreement with
the results of the fits to individual spectra which yield at least two
``break'' energies (at $\sim$ 1.2 and $\sim$ 3.2 keV respectively) 
with spectral indices $\Gamma \sim$
2.1, 2.6 and 2.9.  
This trend was also found (for this same source) by
Giommi et al. (1998), who indeed showed that a good fit can be
obtained using a curved model.

Fitting together the MECS and PDS data yields
spectral parameters very similar to those obtained for the MECS
alone. A cross normalization of 0.8 was used.
The residuals show that the PDS data are consistent with an
extrapolation of the MECS fit up to about 50 keV. Above this energy
the PDS data present a marginal hint of an excess suggesting - as already found for
the PDS alone - a flattening of the spectrum.

Adding also the HPGSPC spectrum to the dataset, and fitting all three
instruments gives again results similar to those obtained for the MECS
alone (Fig. \ref{sei}). Additionally, if one uses MECS data 
above the 3.2 keV break together with HPGSPC data, they are compatible ($\chi^{2}=55$ for 49 DoF)
with a broken power law model with the first spectral index $\Gamma_1$
fixed to the value $\sim$ 2.9 derived from the MECS fit, a break around
10 keV and a flatter slope $\Gamma_2 \sim$ 2.2 (no formal
simultaneous fit of the break energy and $\Gamma_2$ is possible).
This is a further element in favour of a flattening of the spectrum at 
high energies.

\subsubsection{Spectral variability}

In view of a discussion on spectral variability the overall
observation period has been divided into smaller intervals (indicated
as A to F in Fig. \ref{uno}) using the following approximate
intensity windows (referred to a single MECS unit): peak, above 0.8
cts/s (interval A); intermediate (intervals B, D and F); dip, below
0.5 cts/s (intervals C and E). We have then accumulated spectra with
the standard prescriptions described above for each interval, and also
for the combinations B+D+F and C+E. Additionally, we have divided
interval D into two parts (``rise'' and ``fall'') and taken the ratio of
the relevant spectra, which has been found to be consistent with unity.

In order to have a model independent description of the spectral
variability we have computed the spectral ratio between the flaring
(A) and the lowest states (C and E). The result is plotted in
Fig. \ref{sette} ~: the ratio continuously increases with energy,
yielding clear evidence that the flaring state is harder than the low
state.  Since this ratio varies by a factor of about 2 over two
decades the associated change in spectral index can be estimated as
only $\Delta\alpha\simeq 0.15$.

Both the LECS and  MECS spectra for the high, intermediate and low states
 were  separately fitted with broken
power laws, fixing the value of $\Gamma_1$ for the MECS fit at the
value of $\Gamma_2$ obtained from the LECS. 
The results are reported in Table 1. They are consistent
with a hardening of the spectrum with increasing intensity, but the magnitude
of this variation is small, as estimated above, comparable with the errors in the fit
parameters.

\section{Broad Band Spectral Energy Distributions}

In order to estimate the physical parameters of the emitting region we
constructed broad band spectral energy distributions  (SED)
 based  on the deconvolved LECS and MECS $BeppoSAX$ data
at {\it maximum} and {\it minimum} intensity during the present
 observation (Fig. \ref{otto}). At higher energies, the PDS data 
represent averages over the whole $BeppoSAX$ observation period.
At $\gamma$-ray energies  the EGRET spectrum is plotted as
from the discovery observation (Vestrand, Stacy \& Sreekumar
1995) and also with an intensity multiplied by a factor three, to represent
the gamma-ray state of 11-17 November 1997, as communicated by Sreekumar \&
Vestrand (1997). The $BeppoSAX$ observation occurred near the end of
the two-week CGRO observation while the high gamma-ray state was
recorded during the first days. In the absence of a gamma-ray flux 
exactly simultaneous with the $BeppoSAX$ data we
consider the two sets of gamma-ray intensities as encompassing the
actual values.

At UV wavelengths we plot the maximum and minimum fluxes observed with
IUE 
(Edelson et al. 1992, Urry et al. 1993, Pian et al. 1997)
and at other wavelengths the maxima and minima as observed during
the 1991 and 1994 
multiwavelength campaigns (Courvoisier et al. 1995, Pesce et al. 1997).

 The shape of the SED can be interpreted as due to two components: the first
one, peaking in the soft X-ray range, is commonly attributed to synchrotron radiation
while the second, peaking  above 10 GeV as suggested by
the flat $\gamma$-ray spectrum, can be accounted for by inverse Compton scattering
of the synchrotron photons off the high energy electrons that produced them, namely
 the Synchrotron Self--Compton process  (SSC) (e.g. Ulrich, Maraschi \& Urry
1997 and references therein).

A simple version of this model considers emission from a
homogeneous spherical region of radius R, whose motion can be
characterized by a Doppler factor $\delta$, filled with a magnetic
field B and with relativistic particles whose energy distribution is described
by a broken power law (the latter corresponds to 4 parameters: two
indices $n_1$, $n_2$, a break energy $\gamma _bmc^2$ and $K$, a
normalization constant).
As discussed in detail by  Tavecchio, Maraschi \& Ghisellini (1998) 
the seven model parameters listed above can be strongly constrained by using
seven observational quantities, namely the two spectral slopes (in the X- and 
$\gamma$-ray
bands), the frequency and flux of the synchrotron peak, a flux value
for the inverse Compton component emission and a lower limit to the IC
peak frequency. Assuming $R= c t_{var}$ with a variability timescale
$t_{var}=2$ hr, the system is practically closed and we obtain {\it 
univocally} a set of physical parameters for the source, with uncertainties 
depending on those of the observed quantities involved.\\
Fig.4 of Tavecchio, Maraschi \& Ghisellini (1998) shows the allowed region
in the $B-\delta $ space derived for PKS 2155-304 with values of the 
observational quantities encompassing those derived here. A lower limit 
$\delta > 15 $ is set by the "internal" transparency condition for TeV
$\gamma $-rays, while the limits derived from the modeling of the SED fall
somewhat above this.

In Fig. \ref{otto} 
we show two SEDs computed with the SSC models described above aimed
at reproducing the high and low X-ray states observed with $Beppo$SAX. 
We (arbitrarily) assumed that the lower intensity X-ray state  
corresponds to the gamma-ray emission reported in 1995.
The parameters of the model for the lower state have the following values: 
$\delta =18$, $B$=1 G, $R=3\times 10^{15}$ cm,
$n_1=2$, $n_2=4.85$, $K= 10^{4.7}$, $\gamma_b= 10^{4.5}$.
The inverse Compton emission is computed here with
the usual step approximation for the Klein-Nishina cross section,
i.e. $ \sigma = \sigma _T$ for $\gamma \nu _t < mc^2/h$ and $\sigma =0$ 
otherwise, where $\gamma $ is the Lorentz factor of the electron and 
$\nu _t$ is the frequency of the target photon.

The comparison of the flaring spectrum with the lower intensity one
in Fig. \ref{otto}
shows that the {\it peak} in the SED shifted to higher energies during
the flare.  In fact the peak occurs at about 1 keV during the flare,
while in the fainter states it falls towards the lower end of the
$BeppoSAX$ range ($\simeq 0.2$ keV).  A similar behavior was observed
in Mrk 421 with ASCA (Takahashi et al. 1996) and with $BeppoSAX$
(Fossati et al. 1998). A much more extreme case occurred in Mrk 501,
when in a state of exceptional activity the SED peak was observed to
be in the 100 keV range (Pian et al. 1998).

Therefore, in the model  for the flaring state, the break energy of 
the electron spectrum was
shifted to higher energies ($K= 10^{4.8}$, $\gamma_b= 10^{4.65}$) leaving the 
other parameters unchanged.  Correspondingly also the IC peak
increased in flux and moved to higher energies.  Both effects are
however reduced with respect to the ``quadratic'' relation expected in
the Thomson limit since for the required very high energy electrons the
suppression due to the Klein-Nishina regime plays an important role.

The models predict TeV emission at a detectable level. Indeed, towards
the completion of this work, we have been informed of the detection of
high energy $\gamma$--rays by the Mark 6 telescope (Chadwick et
al. 1998). Part of the Mark 6 observing period overlaps with that of
$BeppoSAX$ and EGRET. Although the exact flux level simultaneous to
the $BeppoSAX$ one has not been reported, in November 1997 the source
was seen by the Mark 6 telescope at its highest flux (Chadwick et
al. 1998). The time averaged flux corresponds to $4.2\times 10^{-11}$ ph
 cm$^{-2}$ s$^{-1}$ above 300 GeV (and extending up to $>$ 3 TeV). 

Indeed the model for the lower intensity state
reproduces the average TeV emission flux level remarkably well. A more
detailed test, comparing the TeV fluxes associated with  different 
X-ray states  will be hopefully possible with future observations.

Current models of the IR background (Malkan \& Stecker 1998) predict an
optical depth between 1 and 2 for 1 TeV photon from a source located at 
$z\sim 1$ ( Stecker \& de Jager 1998). The resulting flux reduction would be
still consistent with our model. Signatures of this effect may be found in 
the future from a measurement of the TeV spectral shape.

Note that, given the flat soft X-ray spectrum, it would be impossible
to reproduce with these same models also the non-simultaneous
optical-UV data reported in the figure.  Since the variations occur on
longer time scales in the lower energy bands, the optical-UV flux may
derive from a larger spatial region which acts as a reservoir for the
partially cooled particles.  At higher energies instead we may observe
the radiation from the freshly accelerated particles, before they can
accumulate in the reservoir. In this case, a single homogeneous region
is not sufficient to describe the full broad band SED. 
Alternatively a more complex spectral shape
for the electron energy distribution has to be assumed.

\section{Spectral Dynamics}

The time resolved continuum spectroscopy, made possible by sensitive
and broad band instruments like ASCA and $BeppoSAX$, has triggered the
need for time dependent models describing the changes in particle
spectra due to acceleration, energy losses and diffusion.
The problem is in general complex and
only some simplified cases have been treated up to now (e.g. Kirk, Rieger \&
Mastichiadis 1998, Dermer 1998, Makino 1998). In addition,
light travel time effects through the emitting region may be important
(Chiaberge \& Ghisellini 1998).

A crucial point in this problem is the measurement and interpretation
of lags of the soft photons with respect to the harder ones. These can
be produced through radiative cooling  if the population of
injected (accelerated) electrons has a low energy cut off or possibly
a sharp low energy break, as clearly shown by Kazanas, Titarchuk \& Hua (1998).
If so, the observed lag $\tau _{obs}$ depends only on the
value of the magnetic field (assuming synchrotron losses are dominant, 
as is roughly the case for this source) and $\delta $. Their relation 
can be expressed as

\begin{equation}
B\delta^{1/3} \simeq 300 \left( \frac{1+z}{\nu _{1}} \right)^{1/3}  
\left[ \frac{1-(\nu_1/ \nu_0)^{1/2}}{\tau _{obs}} \right]^{2/3} G
\label{diciassette}
\end{equation}
where $\nu_1$ and $\nu_0$ represent the frequencies (in units of 
$10^{17}$ Hz) at which the
observed lag (in sec) has been measured.

It is interesting to note that the value of the lag (0.5 hrs) inferred from the present
$BeppoSAX$ observations yield a B and $\delta $ combination consistent
with the parameters obtained {\it independently} from the spectral
fitting. This argument  supports  the
radiative interpretation of the observed X--ray variability.   An
alternative possibility recently put forward by Dermer (1998)
is that the flare decay is due to deceleration of the emitting plasma
by entrainment of external matter as proposed for $\gamma$-ray bursts.  
A detailed model would be needed for a quantitative comparison with the
present data.  

We recall that also in Mrk 421, a source closely similar to PKS 2155-304 in 
the SED, similar values of the lag have been found (Takahashi et al. 1996). 
In fact the same arguments
applied here to PKS 2155-304 give similar estimate of the physical parameters
for Mrk 421 (see Tavecchio Maraschi \& Ghisellini 1998). In the case of
the other well established TeV source Mrk 501 lags have not been measured up to
now.

\section{Summary and Conclusions}

Simultaneous multiwavelength monitoring has proved to be 
a very powerful tool to test and constrain emission models for
blazars. The $BeppoSAX$ observations of PKS~2155--304 here reported,
were partially overlapping with an intense $\gamma$--ray flare
detected by EGRET and with observations at TeV energies by the Mark 6 telescope 
which also revealed a high flux.

Although exactly simultaneous $\gamma$-ray data are not available at the time
of writing, the X--ray data alone  already provide us with
interesting results. The X--ray flux was almost at the highest level
ever detected. A curved spectral component, which can be identified
with the high energy end of the synchrotron emission, extends up to about
50 keV, while at higher energies the spectrum flattens, plausibly
revealing the contribution of inverse Compton emission.

Indeed this is what is generally expected in the context of emission from
a relativistic jet, which seems quite convincingly to account for the
SED of blazars.  Within the blazar class, the synchrotron and inverse
Compton components of low power BL Lac objects, as is the case for
PKS~2155--304, tend to peak at the highest energies (X--ray and TeV
energies, respectively) and the synchrotron photons probably dominate the seed
radiation field to be upscattered to $\gamma$--ray energies.

Further constraints on the structure and physical parameters of the
emitting source come from adding to the spectral the temporal
information. Indeed we find that SSC emission from a homogeneous
region can consistently account for both the broad band SED and the
soft lag detected by $BeppoSAX$, the latter being due to radiative
cooling of the high energy part of the electron distribution.

An interesting and testable prediction of this interpretation is the
emission of TeV photons, which indeed has been found {\it a posteriori} to
be in remarkably good agreement with the results recently reported by
the Mark 6 team (Chadwick et al. 1998).

\acknowledgments

We thank L.Piro, $BeppoSAX$ Mission Scientist, and the (chairman and
members of) $BeppoSAX$ Time Allocation Committee for allowing us to
perform this observation in replacement of another target allocated to
us, and the Mission Planning team at $BeppoSAX$ SDC for the prompt
scheduling. We also gratefully acknowledge conversations with D.Dal
Fiume for extremely useful hints on PDS data reduction, and some
suggestions by A.Santangelo about HPGSPC data reduction.  AC and GF
thank the Italian MURST for financial support.

\begin{deluxetable}{lcccccccc} 
\scriptsize
\tablenum{1}
\tablecaption{ Best fit parameters \label{tab1}}
\tablewidth{0pt}
\tablehead{
      \colhead{Data Set}                           &
      \colhead{$N_{H}$       \tablenotemark{a}}    &
      \colhead{$\Gamma$      \tablenotemark{b}}    &
      \colhead{$\chi ^2$/DoF \tablenotemark{b}}    &
      \colhead{$\Gamma _1$   \tablenotemark{c}}    &
      \colhead{$\Gamma _2$   \tablenotemark{c}}    &
      \colhead{$E_b $ (keV)  \tablenotemark{c}}    &
      \colhead{$\chi ^2$/DoF \tablenotemark{c}}    &  
      \colhead{Flux          \tablenotemark{d}}     
}
\startdata
LECS, all      & $2.1_{-0.05}^{+0.1}$ & $2.38 \pm 0.02$ & 207/48     & \nodata                & \nodata                & \nodata       &\nodata    &\nodata \nl
               & $1.4 \pm 0.1$ & \nodata                & \nodata    & $2.09 \pm 0.05$        & $2.54 \pm 0.04$        & $1.1 \pm 0.1$ & 56.4/45   &\nodata \nl
               & 1.36 (fix)    & $2.21 \pm 0.01$        & 57.4/48    & $2.06 \pm 0.02$        & $2.54 \pm 0.04$        & $1.1 \pm 0.1$ & 57.4/47   &\nodata \nl
\tablevspace{10 pt}
MECS, all      & $41 \pm 20$   & $2.92 \pm 0.04$        & 74.1/56    & \nodata                & \nodata                & \nodata       &\nodata    &\nodata \nl
               & 1.36 (fix)    & $2.78 \pm 0.02$        & 118/57     & $2.64_{-0.08}^{+0.06}$ & $2.88_{-0.03}^{+0.07}$ & $3.2 \pm 0.1$ & 76.2/55   &\nodata \nl
\tablevspace{10 pt}
PDS,all        & \nodata       & $2.1_{-0.8}^{+1.7}$         & 5.0/2      & \nodata                & \nodata                & \nodata       &\nodata    &\nodata \nl
\tablevspace{10 pt}
LECS+MECS, all & 1.36 (fix)    & \nodata                & \nodata    & $2.09 \pm 0.02$        & $2.76 \pm 0.02$        & $1.4 \pm 0.1$ & 244.0/105 &\nodata \nl
\tablevspace{10 pt}
MECS+PDS, all & 1.36 (fix)     & \nodata                & \nodata    & $2.65_{-0.09}^{+0.05}$ & $2.90 \pm 0.05$        & $3.2 \pm 0.1$ & 87.2/58   &\nodata \nl
\tablevspace{10 pt}
MECS+HPGSPC+PDS, all & 1.36 (fix)     & \nodata         & \nodata    & $2.65_{-0.08}^{+0.05}$ & $2.89 \pm 0.04$        & $3.2 \pm 0.1$ & 85.8/64   &\nodata \nl
\tablevspace{15 pt}
LECS+MECS \nl
all           & 1.36 (fix)     & \nodata                & \nodata    & $ 2.07 \pm 0.02$       & $2.63 \pm 0.03$        & $1.2 \pm 0.1$   & 153.4/103 & 0.81\nl
              &                &                        &            &   2.63 (fix)           & $2.90 \pm 0.04$        & $3.3 \pm 0.2$   &           & \nl
\tablevspace{5 pt}
A             & 1.36 (fix)     & \nodata                & \nodata    & $ 2.00 \pm 0.04$       & $2.57 \pm 0.07$        & $1.2 \pm 0.2$   & 145.3/102 & 1.27\nl
              &                &                        &            &   2.57 (fix)           & $2.97 \pm 0.12$        & $3.5 \pm 0.5$   &           & \nl
\tablevspace{5 pt}
B+D+F         & 1.36 (fix)     & \nodata                & \nodata    & $ 2.08 \pm 0.02$       & $2.66 \pm 0.03$        & $1.2 \pm 0.1$   & 130.8/103 & 0.85\nl
              &                &                        &            &   2.66 (fix)           & $2.90 \pm 0.05$        & $3.2 \pm 0.3$   &           & \nl
\tablevspace{5 pt}
C+E           & 1.36 (fix)     & \nodata                & \nodata    & $ 2.15 \pm 0.05$       & $2.61 \pm 0.09 $       & $1.1 \pm 0.1$  & 92.7/102  & 0.54\nl          
              &                &                        &            &   2.61 (fix)           & $2.89 \pm 0.07$        & $2.9 \pm 0.2$  &            & \nl
\enddata
\tablenotetext{a}{$10^{20}$ cm$^{-2}$}
\tablenotetext{b}{single power law model}
\tablenotetext{c}{broken power law model}
\tablenotetext{d}{Flux at the source in the 2-10 keV band, in $10^{-10}$ erg cm$^{-2}$ s$^{-1}$}
\tablenotetext{e}{All errors quoted are at 90 \% confidence level}
\end {deluxetable}

\clearpage 

\figcaption[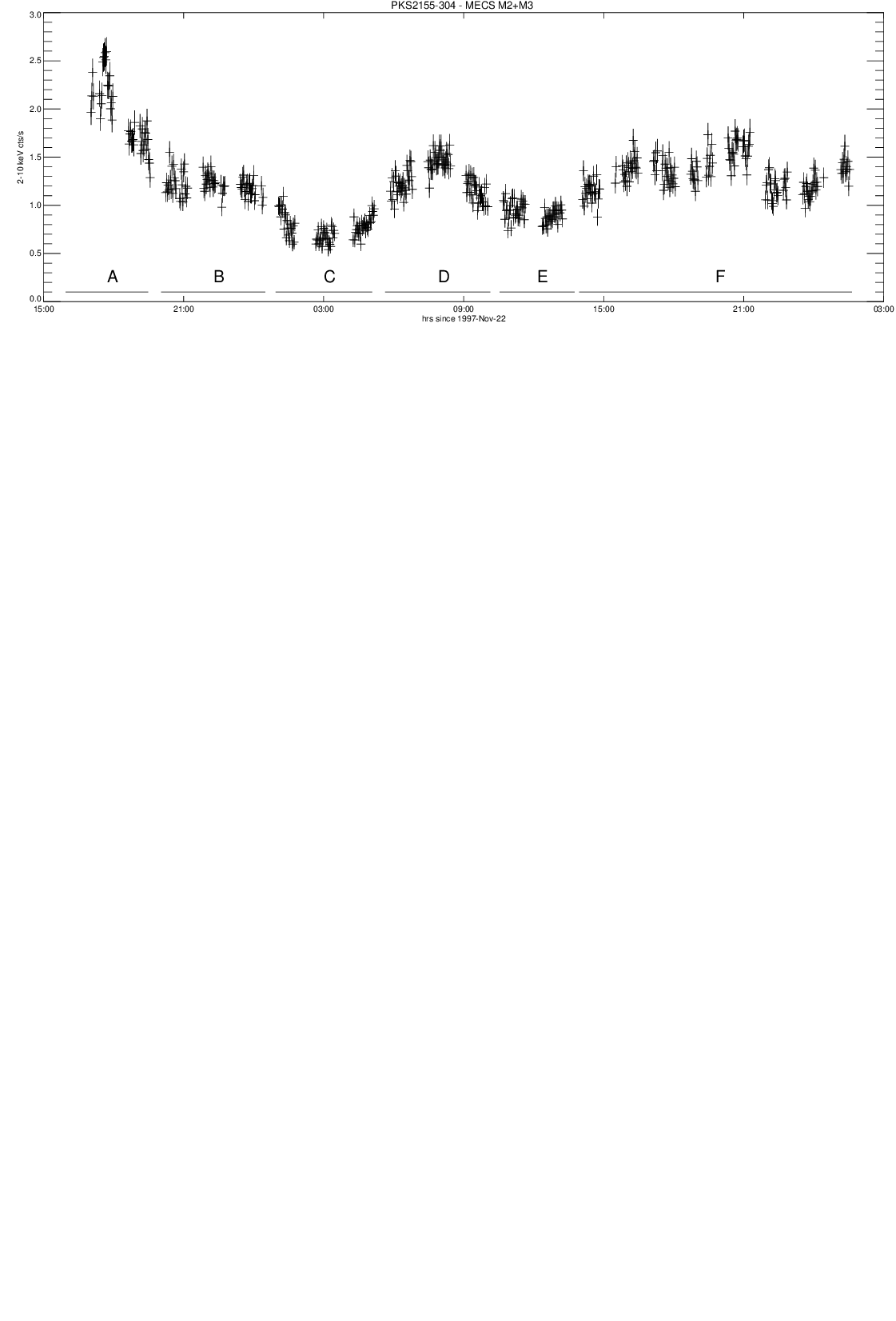]
{MECS light curve at 120 s resolution. As described in the text, the spectral analysis has
also been performed in separate time intervals, indicated as A to F in
the figure. \label{uno} }

\figcaption[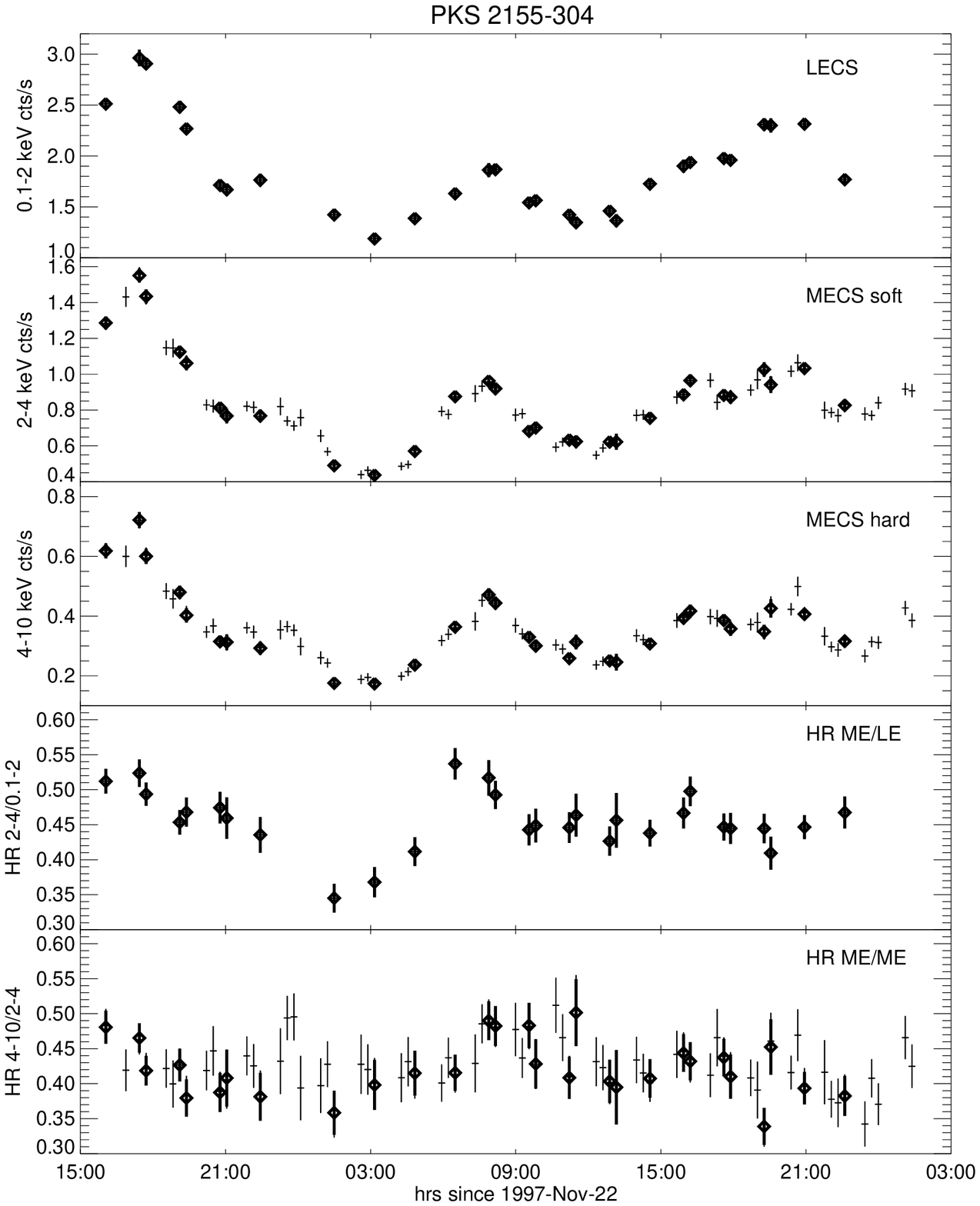]
{Light curves and Hardness Ratios for different energy bands, in 1000 s bins.
In the top three panels we show the light curves in the 0.1-1 keV, 2-4 keV and
4-10 keV energy bands, respectively, while the bottom two panels show the
2-4/0.1-2 and 4-10/2-4 ratios. Note that the coverage of the LECS is more
limited than the one of the MECS, because the LECS is operated only during
dark times, in order to avoid UV light contamination through the entrance
window.  The points simultaneous with the presence of LECS data are indicated
as bold diamonds. The 2-4/0.1-2 Hardness ratio shows a clear correlation with
the source intensity.
\label{due} }

\figcaption[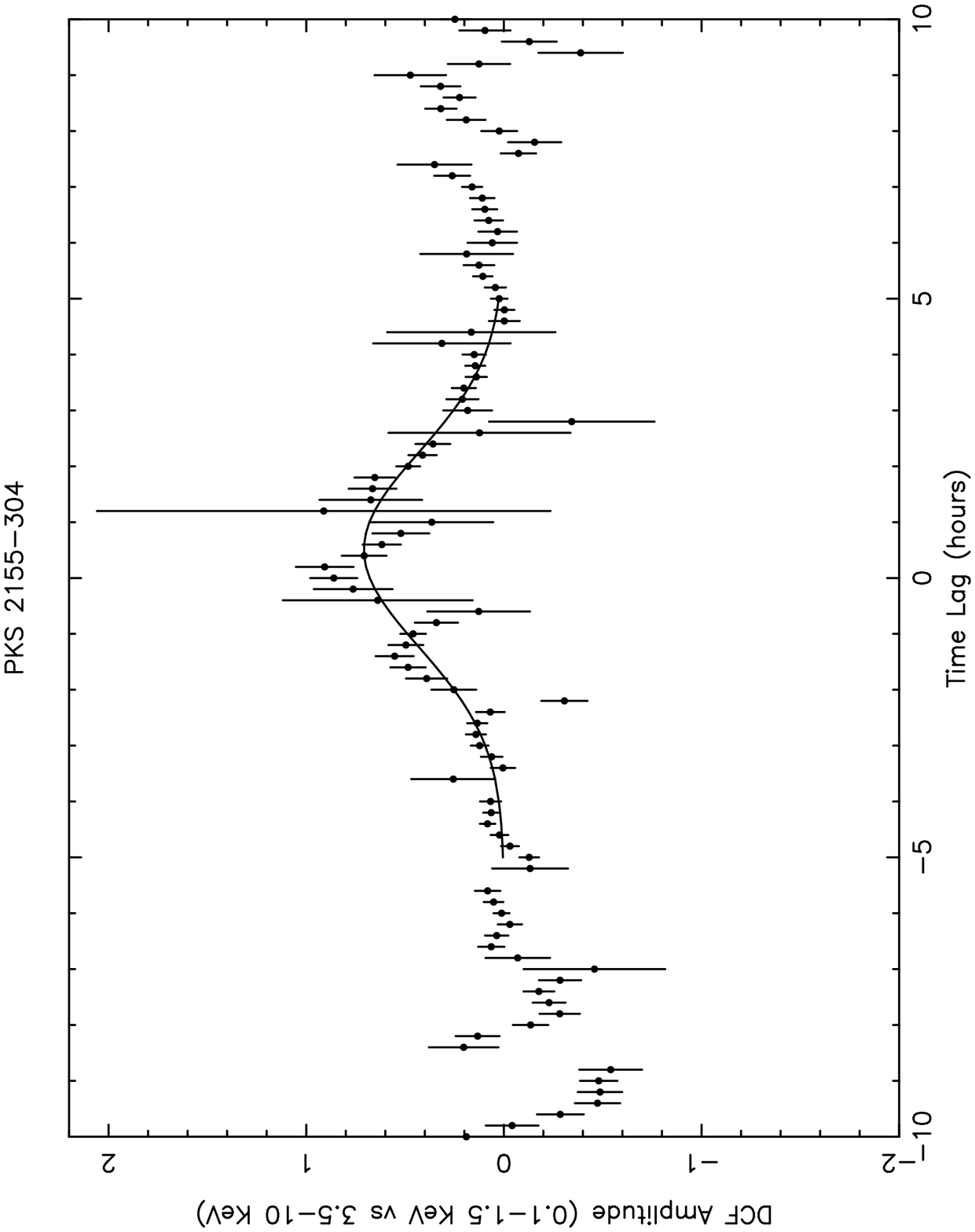]{
Discrete Correlation Function between the 0.1-1.5 keV and 3.5-10 keV
light curves (binned in 300 s intervals, using a DCF bin size of 0.2 hr)
together with its gaussian fit. A positive lag indicates that the high
energy X-rays lead the low energy ones. \label{tre} }

\figcaption[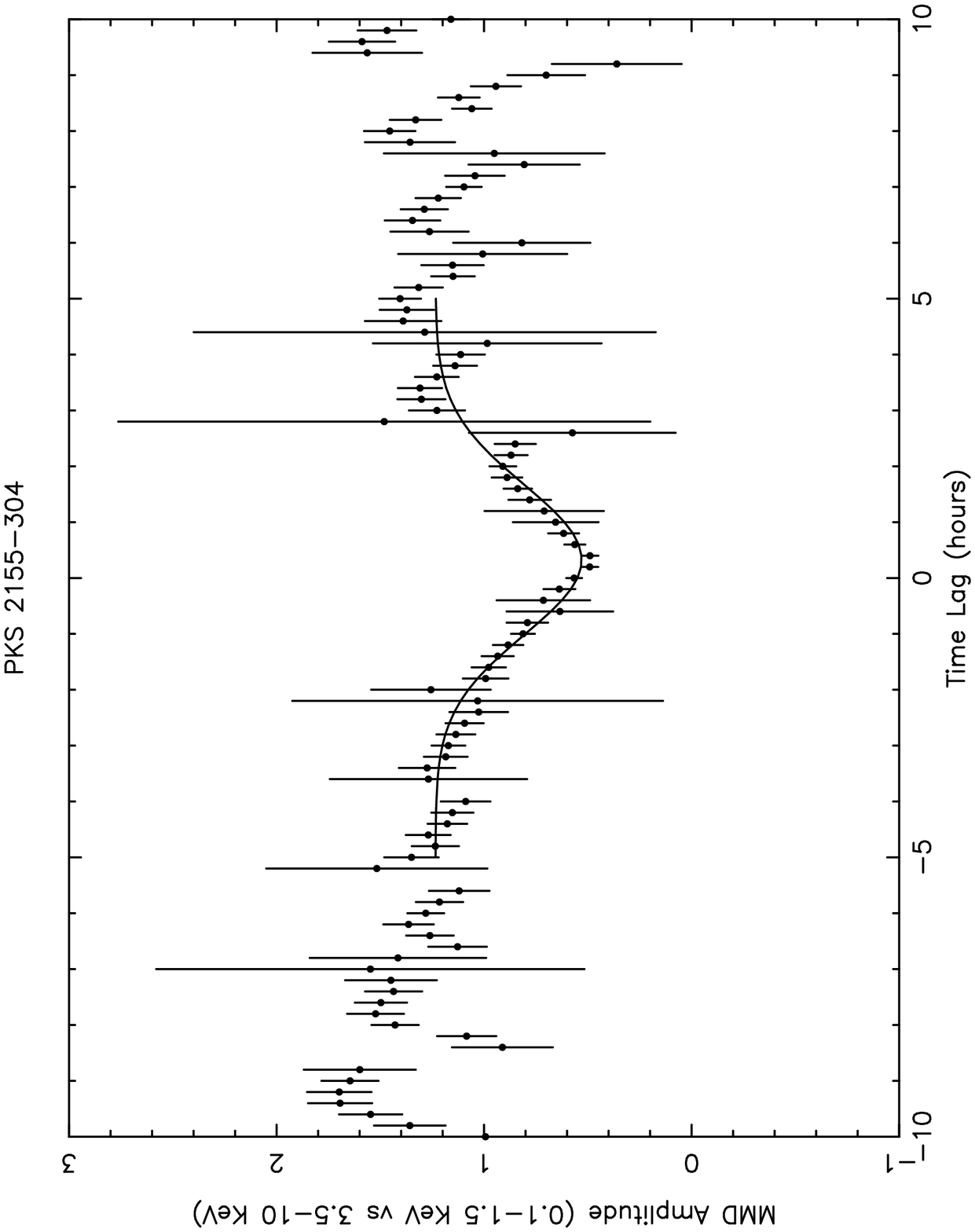]{
MMD for the same light curves and with the same conventions
used in Fig. 3, with its own gaussian fit.  \label{quattro}}

\figcaption[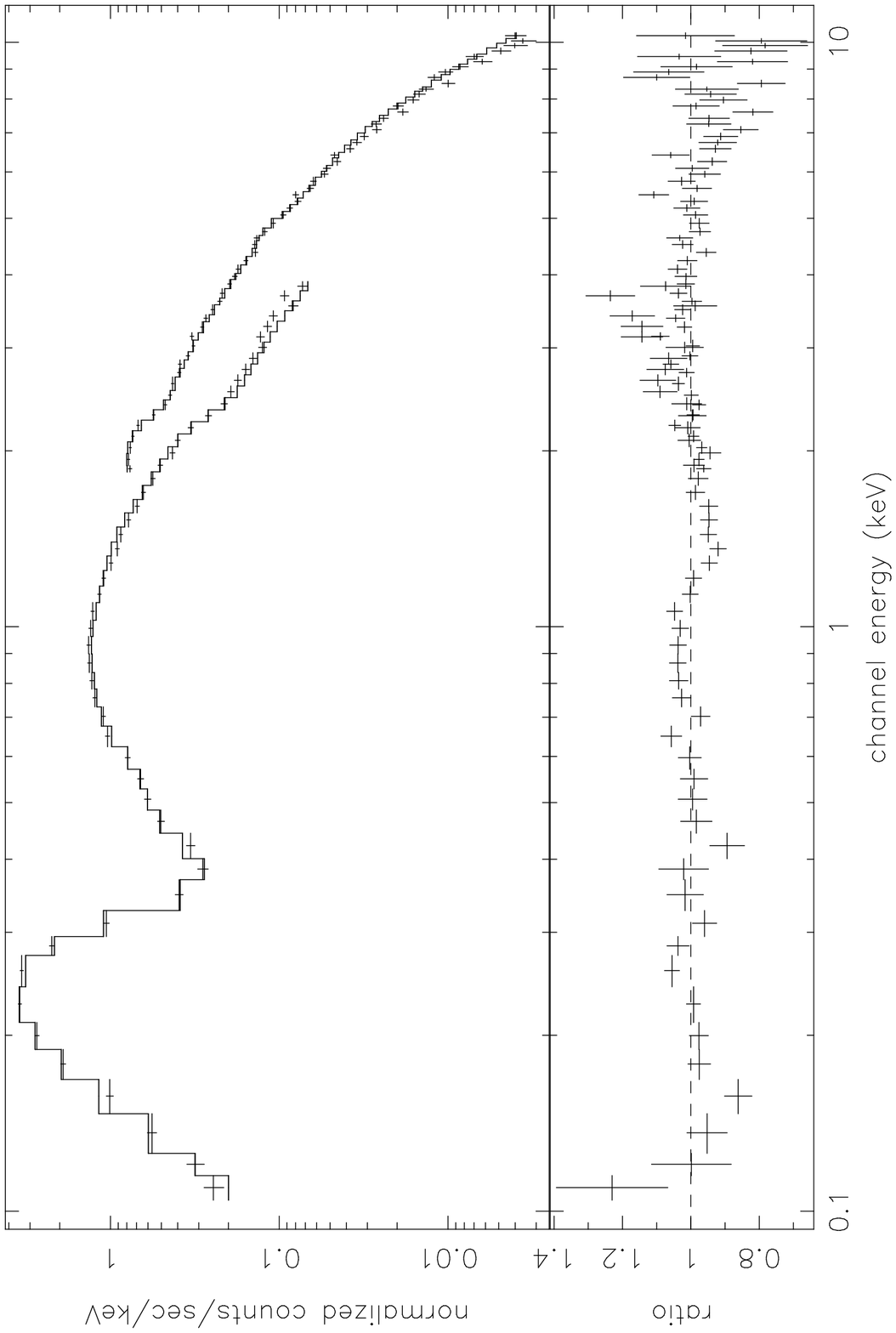]{
LECS+MECS spectra fitted with the broken power law model (see Table
1). \label{cinque} }

\figcaption[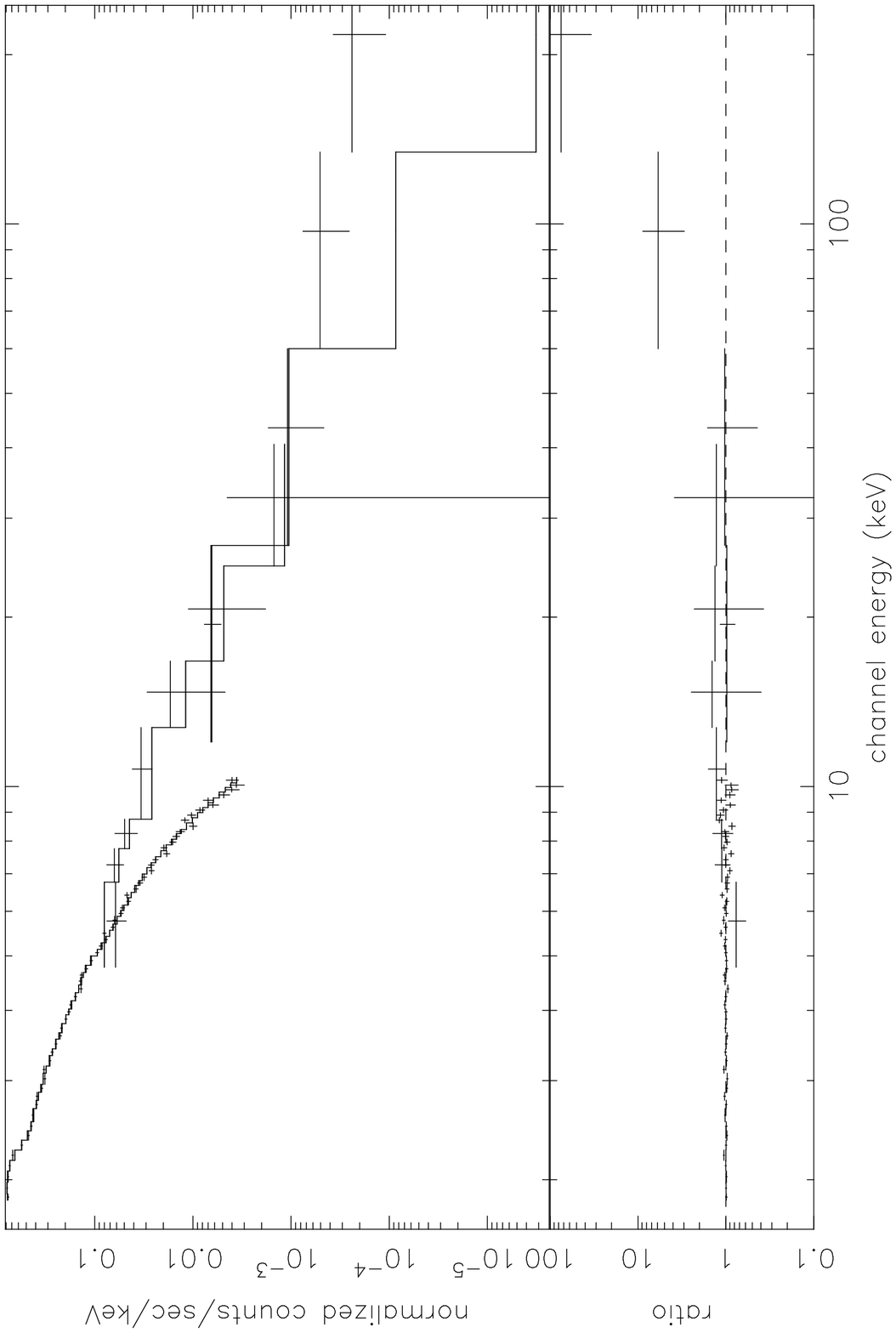]{
MECS+HPGSPC+PDS spectra fitted with a single broken power law. 
Note the logarithmic scale for the data/fit ratio. \label{sei} }

\figcaption[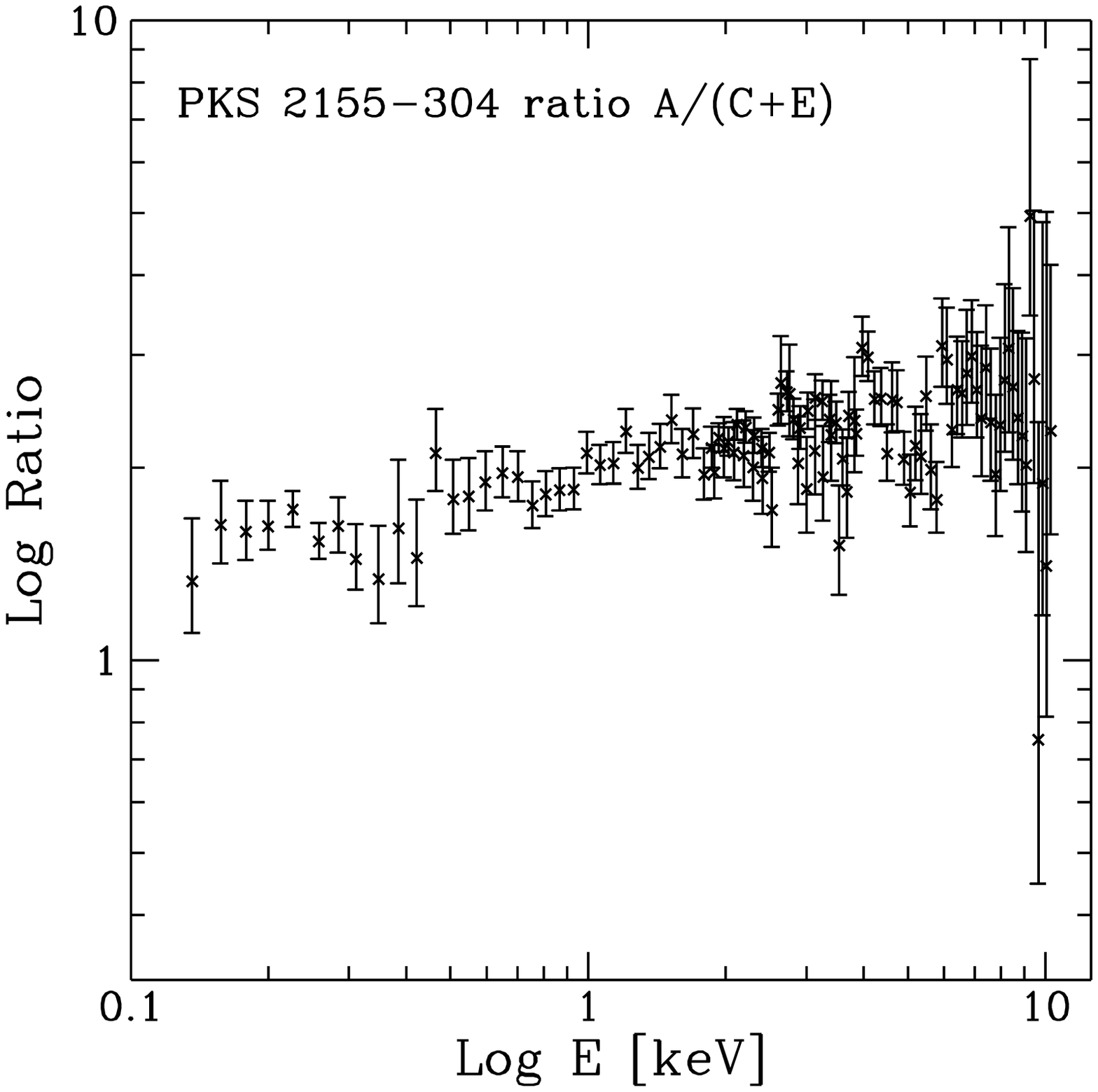]{
Ratio between the flaring state spectrum (A) and the quiescent
spectrum (states C and E). The plot shows that the spectrum hardens
during the flare but the estimated variation in the spectral index is
very small. \label{sette} }

\figcaption[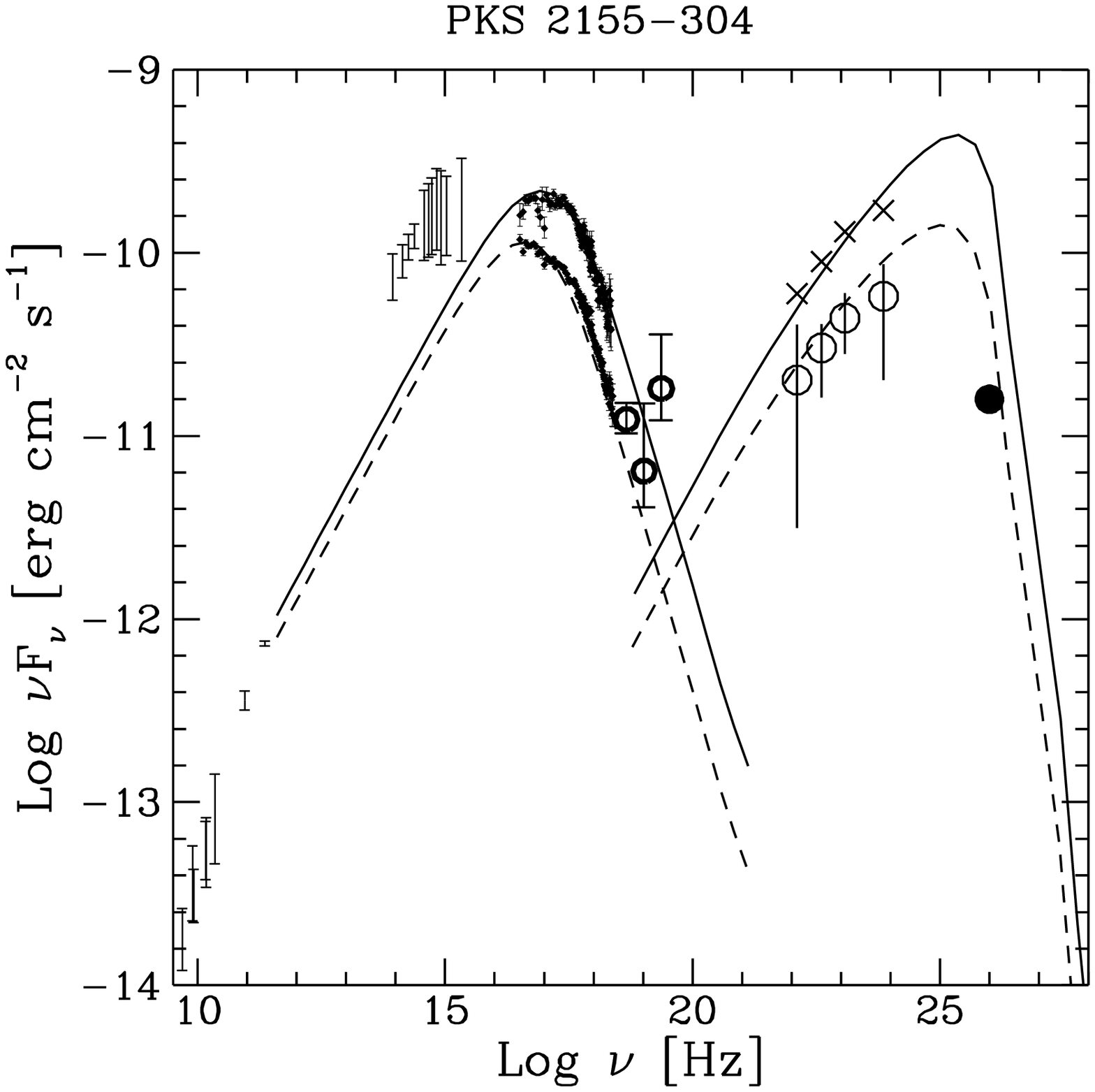]{ The
SED of PKS 2155-304 with the fit obtained with the SSC homogeneous
model in the high (solid) and low (dashed) states.
The parameters of the model are reported in the text.
The X-ray points (error bars) represent the spectra in the high (interval A) and low
(intervals C+E) states.
The thick circles corresponding to the PDS points refer to the average spectrum.
The circles are EGRET $\gamma $-ray data
from Vestrand et al (1995), and they are also shown (as crosses)
multiplied by a factor of three to reproduce the gamma-ray state of
November 1997. The TeV observation (from Chadwick et al. 1998) is also
shown (filled circle).
The vertical bars encompass the range between the minimum and maximum value in
a compilation of radio, optical and UV data (see text for
references).
\label{otto} }

\plotone{figpaper1.ps}

\clearpage 
\plotone{figpaper2.ps}

\clearpage 
\plotone{figpaper3.ps}

\clearpage 
\plotone{figpaper4.ps}

\clearpage 
\plotone{figpaper5.ps}

\clearpage 
\plotone{figpaper6.ps}

\clearpage 
\plotone{figpaper7.ps}

\clearpage 
\plotone{figpaper8.ps}


\clearpage

\begin{thebibliography}{}

\bibitem[Boella et al. 1997a]{1997A&AS..122..299B} Boella, G., Butler,
R. C., Perola, G. C., Piro, L., Scarsi, L., \& Bleeker,
J. A. M. 1997a, A\&AS, 122, 299

\bibitem[Boella et al. 1997b]{1997A&AS..122..327B} Boella, G., et
al. 1997b, A\&AS, 122, 327.

\bibitem[Brinkmann et al. 1994]{1994A&A...288..433B} Brinkmann, W., et al. 
1994, \aap, 288, 433 

\bibitem[Canizares & Kruper 1984]{1984ApJ...278L..99C} Canizares,
C. R. \& Kruper, J. 1984, \apjl , 278, L99

\bibitem[Chadwick P.M. et al. 1998]{dummy} Chadwick, P.M., et al. 1998, 
\apj, accepted (astro-ph/9810209)

\bibitem[Chiaberge & Ghisellini 1998]{dummy} Chiaberge, M. \& Ghisellini, G., 
1998, \mnras, submitted (astro-ph/9810263)

\bibitem[Chiappetti & Torroni 1997]{1997IAUC.6776....2C} Chiappetti,
L. \& Torroni, V. 1997, \iaucirc , 6776, 2

\bibitem[Chiappetti & Dal Fiume 1997]{dummy} Chiappetti, L. \& Dal
Fiume D.1997, Proc. of the 5th Workshop ``Data Analysis in Astronomy''
Erice 27 October - 3 November 1996, Ed. V. Di Ges\'u et al., pag. 101.

\bibitem[Chiappetti et al 1998]{dummy} Chiappetti, L., Cusumano, G.,
Del Sordo, S.,.Maccarone, M.C, Mineo, T. \& Molendi S. 1998,
 Nucl. Phys. B Proc. Suppl., 69, 610

\bibitem[Courvoisier et al. 1995]{1995ApJ...438..108C} Courvoisier, T. 
J.-L., et al. 1995, \apj, 438, 108 

\bibitem[Cusumano et al 1998]{dummy} Cusumano, G., et al. 1998,
Intercalibration of the $BeppoSAX$ Narrow Field Instruments with Crab Nebula,
in preparation

\bibitem[Dermer, Sturner, & Schlickeiser 1997]{1997ApJS..109..103D} Dermer, 
C. D., Sturner, S. J., \& Schlickeiser, R. 1997, \apjs, 109, 103 

\bibitem[Dermer 1998]{1998ApJ...501L.157D} Dermer, C. D. 1998, \apjl, 501, L157 

\bibitem[Edelson et al. 1992]{1992ApJS...83....1E} Edelson, R., Pike, G. 
F., Saken, J. M., Kinney, A., \& Shull, J. M. 1992, \apjs, 83, 1 

\bibitem[Edelson et al. 1995]{1995ApJ...438..120E} Edelson, R., et
al. 1995, \apj , 438, 120

\bibitem[Edelson & Krolik 1988]{1988ApJ...333..646E} Edelson, R. A. \& 
Krolik, J. H. 1988, \apj, 333, 646

\bibitem[Fiore & Guainazzi]{dummy} Fiore, F., \& Guainazzi, M., 1997, 
SAX Scientific Analysis Cookbook : Spectral Analysis,
http://www.sdc.asi.it/software/cookbook/spectral.html

\bibitem[Fossati et al 1998]{dummy} Fossati, G. et al. 1998, 
Nucl. Phys. B Proc. Suppl., 69, 423

\bibitem[Frontera et al. 1997]{1997A&AS..122..357F} Frontera, F.,
Costa,E., Dal Fiume, D., Feroci, M., Nicastro, L., Orlandini, M.,
Palazzi, E., \& Zavattini, G. 1997, A\&AS, 122, 357

\bibitem[Giommi et al. 1998]{1998A&A...333L...5G} Giommi, P., et
al. 1998, \aap, 333, L5

\bibitem[Hufnagel & Bregman 1992]{1992ApJ...386..473H} Hufnagel, B. R. \& 
Bregman, J. N. 1992, \apj, 386, 473 

\bibitem[Lockman & Savage 1995]{1995ApJS...97....1L} Lockman, F. J. \&
Savage, B. D. 1995, \apjs , 97, 1

\bibitem[Kazanas, Titarchuk, & Hua 1998]{1998ApJ...493..708K} Kazanas, D., 
Titarchuk, L. G., \& Hua, X.-M. 1998, \apj, 493, 708 

\bibitem[Kirk, Rieger, & Mastichiadis 1998]{1998A&A...333..452K} Kirk, J. 
G., Rieger, F. M., \& Mastichiadis, A. 1998, \aap, 333, 452 

\bibitem[Madejski et al. 1991]{1991ApJ...370..198M} Madejski, G. M.,
Mushotzky, R. F., Weaver, K. A., Arnaud, K. A., \& Urry, C. M. 1991, \apj , 370, 198

\bibitem[Malkan \& Stecker 1998]{1998ApJ...496...13M} Malkan, M. A. \& 
Stecker, F. W. 1998, \apj, 496, 13 

\bibitem[Manzo et al. 1997]{1997A&AS..122..341M} Manzo, G., Giarrusso,
S., Santangelo, A., Ciralli, F., Fazio, G., Piraino, S., \& Segreto,
A. 1997, A\&AS, 122, 341

\bibitem[Makino et al. 1996]{1996rftu.proc..413M} Makino, F., et al. 1996, 
R\"ongtenstrahlung from The Universe, Ed.  Zimmermann, H.U., Trumper, J.E., Yorke, H.,
MPE Report 263, pag. 413

\bibitem[Makino 1998]{1998bllp.confE.102M} Makino, F. 1998, 
Proc. of the BL Lac Phenomenon meeting, Turku, Finland, 22-26 June 1998, 
Edited by L.O.Takalo, PASP Conf. Series in the press.

\bibitem[Padovani & Giommi 1995]{1995ApJ...444..567P} Padovani, P. \&
Giommi, P. 1995, \apj , 444, 567

\bibitem[Parmar et al. 1997]{1997A&AS..122..309P} Parmar, A. N., et
al. 1997, A\&AS, 122, 309

\bibitem[Pesce et al. 1997]{1997ApJ...486..770P} Pesce, J. E., et al. 1997, 
\apj, 486, 770 

\bibitem[Pian et al. 1997]{1997ApJ...486..784P} Pian, E., et al. 1997, 
\apj, 486, 784 

\bibitem[Pian et al. 1998]{1998ApJ...492L..17P} Pian, E., et al. 1998, 
\apjl, 492, L17 

\bibitem[Sreekumar & Vestrand 1997]{1997IAUC.6774....2S} Sreekumar,
P. \& Vestrand, W. T. 1997, \iaucirc , 6774, 2

\bibitem[Stecker \& De Jager 1998]{1998A&A...334L..85S} Stecker, F. W. \& 
De Jager, O. C. 1998, \aap, 334, L85  

\bibitem[Takahashi et al. 1996]{1996ApJ...470L..89T} Takahashi, T., et al. 
1996,  \apjl, 470, L89 

\bibitem[Tavecchio Maraschi \& Ghisellini 1998]{1998ApJ...509..608T} 
Tavecchio, F. , Maraschi, L.  \& Ghisellini, G.  1998, \apj, 509, 608

\bibitem[Treves 1998]{1998bllp.confE.128T} Treves, A., et al., 1998, 
Proc. of the BL Lac Phenomenon meeting, Turku, Finland, 22-26 June 1998, 
Edited by L.O.Takalo, PASP Conf. Series in the press.
 
\bibitem[Ulrich, Maraschi, \& Urry 1997]{1997ARA&A..35..445U} Ulrich, 
M.-H., Maraschi, L., \& Urry, C.M. 1997, \araa, 35, 445 

\bibitem[Urry & Padovani 1995]{1995PASP..107..803U} Urry, C. M. \&
Padovani, P. 1995, \pasp , 107, 803

\bibitem[Urry et al. 1993]{1993ApJ...411..614U} Urry, C. M., et al. 1993, 
\apj, 411, 614 

\bibitem[Urry et al. 1997]{1997ApJ...486..799U} Urry, C. M., et
al. 1997, \apj , 486, 799

\bibitem[Vestrand, Stacy, & Sreekumar 1995]{1995ApJ...454L..93V}
Vestrand, W. T., Stacy, J. G., \& Sreekumar, P. 1995, \apjl , 454, L93

\bibitem[Zhang et al. 1999]{dummy}
Zhang, Y.H., et al. 1999, \apj , to be submitted

\end{thebibliography}
\end{document}